\documentclass[final,1p,times]{elsarticle}

\usepackage{amssymb}
\usepackage{amsmath}

\journal{Optics Communications}

\begin{document}

\begin{frontmatter}



\title{Self-compression of 5-$\mu$m pulses in hollow waveguides}


\author{Martin Bock$^a$} 
\affiliation{organization={Max Born Institute},
            addressline={Max Born Str. 2a}, 
            city={Berlin},
            postcode={D-12489}, 
      country={Germany}}

\author{Usman Sapaev$^b$} 

\affiliation{organization={Tashkent State Technical University},
            addressline={Universitet St 2}, 
            city={Tashkent},
            postcode={100095},        
            country={Usbekistan}}

            \author{Ji Eun Bae$^c$}

            \affiliation{organization={Centre de Recherche sur les Ions, les Matériaux et la Photonique (CIMAP), UMR 6252, CEA-CNRS-ENSICAEN, Université de Caen Normandie},
            addressline={6 Boulevard Maréchal Juin}, 
            city={Caen Cedex 4},
            postcode={14050},        
            country={France}}

            \author{Anton Husakou$^a$} 

             \author{Joachim Herrmann$^a$}

            \author{Tamas Nagy$^a$}

       \author{Uwe Griebner$^a$}

\begin{abstract}
We experimentally and numerically investigate self-compression of pulses around 5 $\mu$m wavelength in a noble-gas-filled hollow waveguides. We demonstrate spectral broadening of multi-mJ pulses at 4.9 $\mu$m and associated pulse compression from 85 fs to 47 fs in the solitonic pulse compression regime. The self-compression resulted in sub-three-cycle pulses with 17 GW peak power in the 1-kHz pulse train. A numerical model is established and benchmarked against the experimental results. It allows further insights into the pulse compression process, such as scaling of the compression as a function of gas pressure and waveguide radius, and predicts pulse compression in sub-cycle regime for realistic input parameters.
\end{abstract}


\end{frontmatter}




\section{Introduction}
The provision of high-energy few-cycle pulses at even longer wavelength in the mid-infrared (mid-IR) is of important scientific interest. Such pulses are required for investigating key phenomena in strong-field physics \cite{1,2}. The intensity of the driving pulse is essential for achieving higher X-ray flux in the plasma \cite{3}. Due to the wavelength scaling law of high- harmonic generation (HHG), the energy of the generated harmonics can be extended up to the keV range \cite{4}. Furthermore increasing the driver wavelength, the THz conversion efficiency both in the focusing and the filamentation regime can be significantly enhanced \cite{5}.
Unfortunately, only very few sources with peak powers above 1 GW have been demonstrated for wavelength beyond 4 µm \cite{6}. These are typically based on optical parametric chirped pulse amplification (OPCPA). With regard to the applications mentioned above, operation with repetition rates in the kHz range is desired to ensure a sufficient signal-to-noise ratio for the measurement.
The OPCPA system generating the highest peak power beyond 4 $\mu$m of 33 GW delivered 89 fs pulses with 3.4 mJ at 5 µm and operated at 1 kHz repetition rate \cite{7}. Another OPCPA equipped with five OPA stages operated at 7 µm wavelength and delivered 188 fs pulses with energy of 0.75 mJ however, at a lower repetition rate of 100 Hz \cite{8}. Both systems contain ZnGeP$_2$ as nonlinear crystals and are pumped at 2 µm wavelength.
The pulse durations achieved are not yet in the few-cycle regime, necessary to efficiently drive processes in strong-field science. For entering this regime, spectral broadening of the pulses by means of self-phase modulation (SPM) in a nonlinear medium with subsequent re-compression is a common method. This technique of nonlinear post-compression was successfully applied in the near infrared by employing hollow-core fibers (HCFs) \cite{9} or multi-pass cells \cite{10}. For the latter no reports exist for wavelengths beyond 4 µm.

Hollow-core fibers are an ideal fiber type for the distortionless delivery of ultrashort high peak power pulses. If the power of the input pulses is higher than the threshold for the formation of a fundamental soliton the input pulses experience an initial temporal compression and subsequently breaks into several fundamental solitons accompanied with the emission of a supercontinuum.

With regard to HCF compression at longer wavelength, recently studies have been published at 4.0 µm wavelength. 105 fs pulses of a KTA based OPCPA with 5 mJ energy were sent through a Kr-filled HCF to yield 2.6-mJ, 21.5-fs pulses at 100 Hz repetition rate \cite{11}. Using coherent beam combining of two similar OPCPAs but at a lower repetition rate of 20 Hz, 160-fs, 5.6-mJ pulses were injected in a single Kr-filled HCF delivering post-compressed pulses with 2.7 mJ energy and 22.9 fs duration \cite{12}. With the availability of novel pump sources a high-energy Fe:ZnSe CPA at 400 Hz repetition rate was realized at 4.0 µm. It emits 238 fs pulses with 2.8 mJ energy. Also using a Kr-filled HCF, the pulses were post-compressed to 42 fs with 1 mJ energy \cite{13}.
Here, we investigate the potential of the HCF for the compression of mid-IR pulses at even longer 5-µm wavelength due to solitonic effects. We experimentally show that significant spectral broadening can be achieved in an argon-filled HCF, accompanied by the pulse compression. We establish a numerical model for the pulse propagation simulation and benchmark it against the experiment. Using this model we analyze the scaling of the compression factor and predict that compression of infrared pulses down to single-cycle durations is possible. 

The paper is organized as follows: In Section 2, we present the experimental setup and results, as well as numerical model and its benchmarking. Section 3 is devoted to the scaling of the compression factor and the numerical prediction of the single-cycle compression, followed by conclusion.

\section{Experimental results}
\label{sec1}

\begin{figure}[ht!]
\centering
\includegraphics[width=1.0\textwidth]{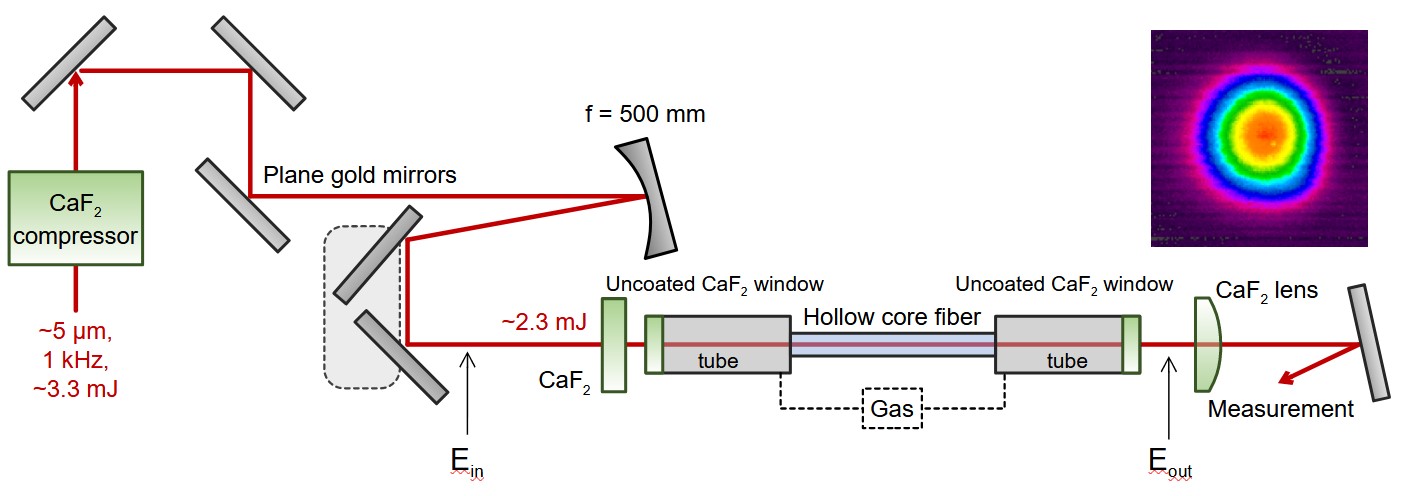}
\caption{Experimental setup. 5-$\mu$m idler pulses of the OPCPA are compressed in CaF$_2$ bulk material and focused onto the HCF entrance by a reflective telescope (M2, $f$ = 500 mm, M3, $f$ = -4000 mm) after passing through a 1.5-mm-thick uncoated CaF$_2$ window. The HCF is placed in a gas chamber with controllable argon pressure up to 3.6 bar. The output pulses are characterized by a scanning monochromator and/or a home-made SHG-FROG setup. Inset: Far-field intensity distribution of the self-compressed pulse (FOV: field-of-view). }\label{fig1}
\end{figure}

The experimental setup is depicted in Fig. 1 in a simplified form. The 5-µm input pulses are provided by a mid-IR OPCPA operating at a repetition rate of 1 kHz. The system, described in \cite{7}, is based on a multi-stage ZnGeP$_2$ parametric amplifier delivering idler pulses with pulse energy around 3 mJ. The pulses are compressed in a CaF$_2$ bulk compressor consisting of four rods at Brewster’s angle, an AR-coated window and an uncoated collimating lens. The spectrum covers the width from 4.3 $\mu$m to 5.6 $\mu$m (at 10$^{-2}$ level), with a smooth, nearly constant spectral phase, only sinusoidal oscillations are visible, which are caused by the GTI effect of the pulse shaper in the OPCPA and a small residual amount of positive third-order dispersion [Fig. 2(a)]. As a result, the retrieved pulse FWHM duration of 85 fs is close to the duration of the Fourier-transform-limited (FTL) pulse FWHM of 68 fs, as shown in Fig. 2(b). The above results are based on the measured [Fig. 2(c)] and retrieved [Fig. 2(d)] SHG-FROG traces.

After compression the 5-$\mu$m input beam is focused by an astigmatism-compensated mirror telescope (M2, M3). Its geometry is adjusted so that the maximum matching with the fundamental mode of the HCF is achieved. Due to the loss during propagation, 1.9 mJ pulse energy is available for coupling into the fiber entrance. Both ends of the gas container are equipped with uncoated 1.5-mm thick CaF$_2$ windows. We used fibers with 50 cm length, filled with argon up to 3.6 bar, and inner radius of 250 $\mu$m, as a trade-off between the transmission of the waveguide (exhibiting high linear losses in the IR) and the amount of the spectral broadening, which is proportional to the length. The pulses that have passed through the HCF were directed to a monochromator (Horiba Jobin Yvon iHR320) equipped with a mercury cadmium tellurid detector, in addition, their temporal profiles were measured by a home-built second-harmonic generation frequency-resolved optical grating (SHG-FROG) setup. During propagation of the beam, the energy remains in the fundamental mode of the fiber yielding excellent beam quality typical for HCFs, as illustrated by the far-field intensity distribution in the inset of Fig. 1.

\begin{figure}[ht!]
\centering
\includegraphics[width=1.0\textwidth]{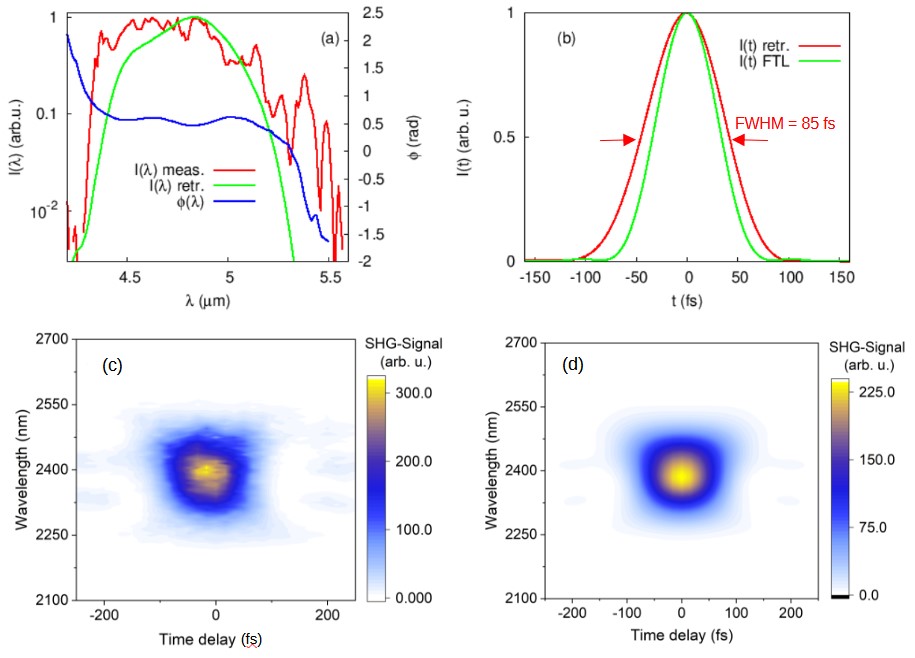}
\caption{Input pulse characterization. In (a), measured and retrieved spectra are shown by red and green curves, correspondingly, together with the retrieved spectral phase (blue curve). In (b), retrieved intensity shape (red curve) is shown together with Fourier-transform-limited (FTL) intensity shape (green curve). In (c) and (d), measured and retrieved SHG-FROG traces are depicted. }\label{fig2}
\end{figure}

\begin{figure}[ht!]
\centering
\includegraphics[width=1.0\textwidth]{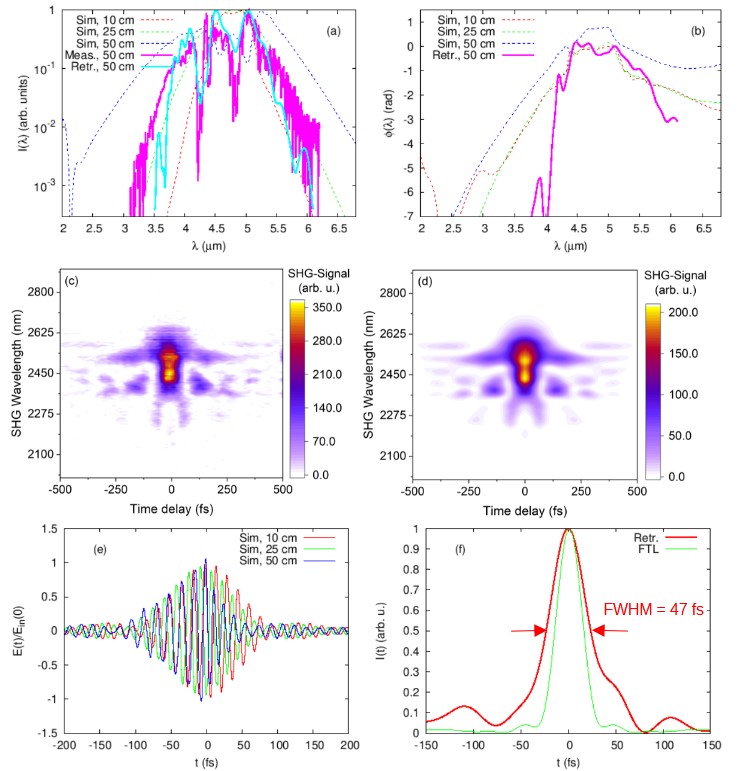}
\caption{Experimental results of HCF self-compression and benchmarking of the numerical model. In (a), measured (magenta) and retrieved (light blue) experimental spectra are shown together with numerically predicted spectra depicted by red, green, and blue dashed curves after 10 cm, 25 cm, and 50 cm of propagation, correspondingly. In (b), retrieved (magenta) experimental spectral phases are shown together with numerically predicted spectral phases depicted by red, green, and blue dashed curves after 10 cm, 25 cm, and 50 cm of propagation, correspondingly. The above measurement are based on the measured [Fig. 3(c)] and retrieved [Fig. 3(d)] FROG traces. In (e), the temporal profiles of the electric field are presented by red, green, and blue curves after 10 cm, 25 cm, and 50 cm of propagation, correspondingly. In (f), retrieved (red curve) intensity profile is presented, together with the FTL intensity profile (thin green curve). A 50-cm-long fiber with inner diameter of 500 $\mu$ m, homogeneously filled with 3.6-bar argon gas, was used. Input pulse has the energy of 1.9 mJ and FWHM duration of 85 fs, with spectrum centered around 4.9 $\mu$m, as shown in Fig. 2. }\label{fig3}
\end{figure}

For our operation regime, we have not observed any fiber damage and no heating effects by the nearly 2 W average power. The measured overall throughput of the HCF unit was 50\% which is slightly lower than the theoretically calculated 61\% taking the linear fiber losses and the Fresnel reflection losses of the uncoated windows into account. We attribute the difference to sub-optimal coupling into the HCF as well as non-perfect straightness of the waveguide. The experimental and retrieved spectra after SPM broadening in the HCF, as shown in Fig. 3(a) by the magenta and light blue curves, respectively, extends from 3.5 $\mu$m to almost 6 $\mu$m which is significantly larger than the spectral width of the input pulse. The spectra are measured after 3 m propagation in air. Therefore, the recorded SPM broadened spectra are additionally modulated by the strong CO$_2$ absorption band at 4.2 $\mu$m and the absorption lines of water vapor in the atmosphere above 5.3 $\mu$m. Only reflective optics are used to propagate the self-compressed pulse except a ZnSe-lens (GDD: $\sim$0 fs$^2$ at 5 $\mu$m) is used in the FROG-setup. A 100-$\mu$m thick GaSe crystal is implemented as nonlinear crystal in the FROG which limits the phase-matching bandwidth, i.e., spectral components below 3.5 $\mu$m do not contribute to the pulse characterization. Spectral parts of the self-compressed pulse above 5.3 $\mu$m are strongly influenced by the water-vapor absorption because the experimental setup is not purged. The reconstructed spectral phase, indicated in Fig. 3(b) by the magenta curve, does not show a large variation in the range from 4 $\mu$m to 5.5 $\mu$m, which, together with the expected anomalous dispersion of the fiber, suggests the possibility that the output pulse is self-compressed with duration close to the duration of the FTL pulse. Indeed, as shown in Fig. 3(f), the retrieved self-compressed pulse duration from the SHG-FROG measurement is 47 fs. Thus the compression factor is 1.8. The self-compressed pulse duration approaches the FHWM of the FTL pulse of 31 fs, shown in Fig. 3(f) by the thin green curve and corresponds to sub-three optical cycles. For the measured pulse energy of 0.88 mJ after the CaF$_2$ exit window of the gas container, the 47 fs pulse duration translates into a remarkable peak power of 17 GW. The phase retrieval is based on the measured [Fig. 3(c)] and retrieved [Fig. 3(d)] SHG-FROG traces and performed on a square grid of 256 pixel size. The FROG error amounts to 0.8\%. 

To further investigate the spectral broadening near 5 $\mu$m, we numerically simulate pulse propagation using the forward Maxwell equations \cite{hh}:

\begin{equation}
    \frac{\partial E(z,\omega)}{\partial z}=i\frac{\omega}c[n(\omega)-n_g]E(z,\omega)-\frac{i\omega}{2cn(\omega)}P_{NL}(z,\omega)
\end{equation}
where $E(z,\omega)$ is the Fourier transform of the electric field, $z$ is the propagation coordinate, $n(\omega)$ is the complex-valued effective refractive index which includes the waveguide group-velocity dispersion (GVD) and wavelength-dependent loss, $n_g$ is the group refractive index which characterized the moving coordinate frame, and $P_{NL}(z,\omega)$ is the Fourier transform of the nonlinear polarization. This formalism does not rely on the slowly-varying envelope approximation and can treat very broad spectra, which will be important for this study. Since the critical power of self-focusing is not reached, and the input light couples almost exclusively into the fundamental mode, we neglect coupling to higher-order modes. We calculate $n(z,\omega)$ using Sellmeyer-type expression for the argon refractive index \cite{sell}, and include waveguide dispersion and loss using the well-known Marcatili formalism \cite{marc}. Intensities below the photoionization threshold are considered, therefore the nonlinear polarization is limited to the Kerr term $P_{NL}=p\epsilon_0\chi_3E^3(z,t)$, where $\chi_3=3.54\times10^{-26}$ m$^2$/V$^2$/bar is the third-order nonlinear susceptibility which we assume frequency-independent and $p$ is pressure. Propagation in air which occurs after the fiber in the experimental setup was not included in the simulation.

The results of the simulation are shown in Fig. 3. The numerical spectra, shown in Fig. 3(a) by red, green and blue dashed curves for the consecutive propagation lengths as given in the caption, show a somewhat larger width than the experimental one. In fact, the experimental spectrum better coincides with the numerical spectrum at 25 cm rather than with that at 50 cm. We attribute this to the underestimated loss in the numerical model; in experiment, we expect additional loss both at the fiber entrance and during propagation. The simulated phases, as shown in Fig. 3(b), exhibit a good agreement with the experimentally retrieved phase, which suggests that the experimentally observed pulse compression is predicted also in the numerical simulation. Indeed, the simulated pulse shapes, as presented in Fig. 3(e), indicate compression from input duration of around 85 fs to around 40 fs. Note that we experimentally obtain a shortening from 80 fs to 47 fs, which is somewhat less than  the numerically predicted shortening to 40 fs. We again attribute this discrepancy to the additional experimental losses which were not included in the model.

\section{Discussion}

\begin{figure}[ht!]
\centering
\includegraphics[width=0.7\textwidth]{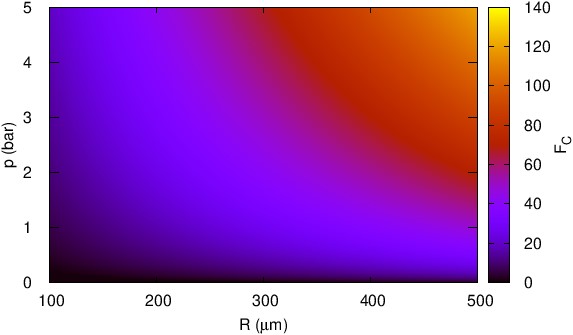}
\caption{Analytical pulse compression factor. Input pulses with FWHM of 85 fs and intensity of 25 TW/cm$^2$, corresponding to the experimental situation of Figs. 2 and 3, were considered.}\label{fig4}
\end{figure}

Let us investigate the mechanism and the potential of the 5-$\mu$m pulse self-compression in the hollow fibers. In contrast to the visible and near-IR regimes, the input spectrum of the 5-$\mu$m pulse lies deeply in the anomalous regime for all the realistic gas pressures and fiber diameters, since the anomalous GVD originating from the waveguide contribution dominates the normal GVD which comes from the argon gas filling. This means that the soliton self-compression will govern the pulse dynamics, with generation of non-solitonic radiation and soliton fission playing only a minor role. The soliton self-compression is determined by the soliton number $N$ of the input pulse, with compression factor $F_C$ (defined as ratio of the input pulse duration to the compressed pulse duration) being given by $F_C\sim 4N$ \cite{argawal}. In the cases when the propagation length required to obtain the shortening was larger than the effective propagation length as determined by loss, we multiplied the compression factor by the ratio of the effective and the required propagation lengths. In Fig. 4, we plot the compression factor $F_C$ as a function of the gas pressure and the waveguide radius, for input pulse duration of 85 fs and constant input intensity of 25-TW/cm$^2$. Note that keeping pulse intensity constant for increasing waveguide radii requires higher pulse energy. One can see that for the typical parameter ranges such as pressures around 2 bar and waveguide radii around 200 $\mu$m, one can expect compression factors of the order of 10. For the considered input pulses this corresponds to single-cycle durations or ever below, and intensity increase up to 150 TW/cm$^2$. Further compression would be impossible, both from the point of view of the waveform which cannot get significantly shorter than single-cycle, and from the point of view of intensity, since photoionization-induced clamping limits the peak intensity to around 100-200 TW/cm$^2$. Therefore the compression regimes with $F_C$ above $\sim$15, as shown in Fig. 4 for large pressures and radii, are physically not accessible.

\begin{figure}[ht!]
\centering
\includegraphics[width=0.7\textwidth]{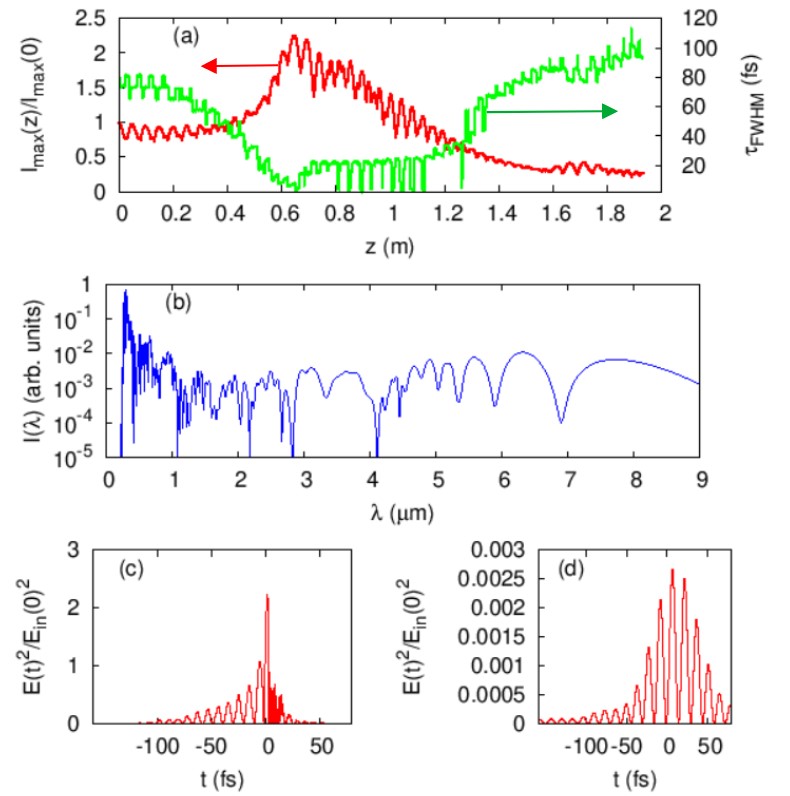}
\caption{Numerical results of pulse compression for a 2-m-long fiber. In (a), the ratio of the $z$-dependent peak intensity to the input peak intensity is shown by red curve, while the dynamics of the FWHM duration is depicted by the green curve. The simulated spectrum and temporal profile after 0.67 m of propagation are shown in (b) and (c), correspondingly. In (d), the temporal profile of the spectral content between 3.5 $\mu$m and 6 $\mu$m is shown. A 2-m-long fiber with inner diameter of 500 $\mu$ m, homogeneously filled with 3.6-bar argon gas, was considered. Input pulse has the energy of 2.1 mJ and FWHM duration of 85 fs, with spectrum centered around 4.9 $\mu$m. }\label{fig5}
\end{figure}

In Fig. 5 we show the results of the numerical simulation of the pulse compression in the regime of large compression factors. A 2-m-long fiber fiber with inner diameter of 500 $\mu$m, filled with 3.6-bar argon gas, was considered. Input pulse has the energy of 2.1 mJ and FWHM duration of 85 fs, with spectrum centered around 4.9 $\mu$m. The FWHM pulse duration, as shown in Fig. 5(a) by the green curve, reduces to below 10 fs with propagation, which corresponds to values below a single-cycle duration. The oscillatory behavior of the FHWM originates from the discrete half-cycles changing their peak intensity with propagation from just above half-maximum of the peak intensity to just below half-maximum, and vice versa. The peak intensity of the pulse, as shown in Fig. 5(a) by the red curve, reaches values more than twice the input peak intensity, before starting to reduce. This intensity behavior is a characteristic signature of the solitonic compression mechanisms. The oscillations of the intensity ratio is caused by carrier-envelope-offset of the pulse changing with propagation, due to difference between the group and phase refractive indices.

In Fig. 5(b) the spectrum of the pulse after 0.67 m of propagation, at the point of maximum compression, is shown. One can see a broad supercontinuum extending from 0.2 $\mu$m to above 8 $\mu$m (width of 5 octaves), which is sufficient to support a generation of single-cycle pulses, with a pronounced peak around 300 nm which we identify as non-solitonic radiation (NSR) \cite{hh}. In Fig. 5(c), we depict the temporal profile of the pulse after 0.67 m of propagation. One can see that while there is a single sharp peak at the center of the pulse, which yields the FHWM of around 3 fs, there is also very significant pedestal. In this sense FWHM measure of the pulse duration underestimates the real pulse duration in our case. Also, after the center of the pulse, we predict very fast oscillations of the electric field. These oscillations are not a numerical artifact; they correspond to the generation of the NSR at wavelength around 300 nm, as shown in Fig. 5(b). 

The presence of the radiation in the near-IR, visible, and even ultraviolet spectral ranges significantly influences the simulated pulse duration. In experiment, however, only the spectral range from roughly 3.5 $\mu$m to 6 $\mu$m contributes to the measured pulse, due to limitations of the characterization system. To illustrate the effect of these limitations, in Fig. 5(d) we show the temporal profile of the simulated pulse whereby we only retain the radiation in the spectral range from 3.5 $\mu$m to 6 $\mu$m. One can see that the pulse becomes significantly longer with a FWHM of 33 fs, which is not far from the experimental value of 47 fs achieved after 0.5 m of propagation.

\section{Conclusion}
In conclusion, we have demonstrated and characterized the self-compression of pulses at 5 $\mu$m wavelength, based on propagation in an argon-filled hollow-core fiber (HCF). 1.9 mJ pulses in a 1-kHz pulse train experienced a 1.8-fold compression, from 85 fs to 47 fs, which corresponds to only sub-three optical cycles. The compressed pulses contained 0.88 mJ energy resulting in a remarkable peak power of 17 GW.

The numerical simulations, in good agreement with the experiment, indicate that the compression is attributed to the solitonic self-compression regime. We numerically investigate the theoretical perspectives of solitonic self-compression in the infrared regime, predicting the possibility to generate single-cycle pulses.

The pulse characterization was affected by absorption in air and it was not possible in the full spectral range of the pulse, mainly because of the limited phase-matching bandwidth of the used nonlinear crystal (GaSe) in our SHG-FROG setup. To characterize the pulses with their full bandwidth, other techniques are required, such as electro-optic sampling. Implementing the latter and evacuating the beam path including the measurement equipment will be the next step.  

Further possibilities to optimize the compression include tailoring of the temporal and spectral profiles of the input pulse, using coated fibers to reduce the linear losses, as well as careful choice of the fiber diameter and gas pressure to minimize the effects of third-order dispersion and other perturbing effects. These directions will be the subject of subsequent investigations.



\section{Acknowledgements}

Authors acknowledge support from the German Research Council (DFG), project numbers GR 2115/6-1, HE 2083/24-1, and HU 1593/11-1, and H2020 Marie Sklodowska-Curie Actions (101034329) Region Normandie, France (Programme WINNINGNormandy)

\section{Data availability}
Data will be made available on request.




\end{document}